# Connecting Beliefs, Mindsets, Anxiety, and Self-Efficacy in Computer Science Learning: An Instrument for Capturing Secondary School Students' Self-Beliefs

Luis Morales-Navarro, *University of Pennsylvania, luismn@upenn.edu*
Michael T. Giang, *California State Polytechnic University, Pomona, mtgiang@cpp.edu*
Deborah A. Fields, *Utah State University, deborah.fields@usu.edu*
Yasmin B. Kafai, *University of Pennsylvania, kafai@upenn.edu*

**Background and Context**: Few instruments exist to measure students' CS engagement and learning especially in areas where coding happens with creative, project-based learning and in regard to students' self-beliefs about computing.
**Objective**: We introduce the CS Interests and Beliefs Inventory (CSIBI), an instrument designed for novice secondary students learning by designing projects (particularly with physical computing). The inventory contains subscales on beliefs on problem solving competency, fascination in design, value of CS, creative expression, and beliefs about context-specific CS abilities alongside programming mindsets and outcomes. We explain the creation of the instrument and attend to the role of mindsets as mediators of self-beliefs and how CSIBI may be adapted to other K-12 project-based learning settings.
**Method**: We administered the instrument to 303 novice CS secondary students who largely came from historically marginalized backgrounds (gender, ethnicity, and socioeconomic status). We assessed the nine-factor structure for the 32-item instrument using confirmatory factor analysis and tested the hypothesized model of mindsets as mediators with structural equation modeling.
**Findings**: We confirmed the nine factor structure of CSIBI and found significant positive correlations across factors. The structural model results showed that problem solving competency beliefs and CS creative expression promoted programming growth mindset, which subsequently fostered students' programming self-concept.
**Implications**: We validated an instrument to measure secondary students' self-beliefs in CS that fills several gaps in K-12 CS measurement tools by focusing on contexts of learning by designing. CSIBI can be easily adapted to other learning by designing computing education contexts.

**Introduction**
The development and use of standardized measurements are needed to improve the rigor and validity of research on students' engagement with computing as it is becoming a standard subject in K-12 education (e.g., Margulieux et al., 2019). Developing standardized instruments will allow for the comparison of outcomes across studies and contribute to building an evidence-based research agenda in K-12 computing education. This approach not only applies to measurements of students' learning of computational concepts and skills but also to non-cognitive factors —beliefs, behaviors, attitudes, habits and dispositions (McGill et al., 2019)—impacting their computing learning experiences. The latter is the focus of this paper. In particular, students' self-beliefs about computing play an important role in supporting—or hindering—the learning of computing (Malmi et al., 2020).

In a recent systematic literature review, Margulieux and colleagues (2019) noted several challenges for improving standardized assessments in computing education. After reviewing close to 200 studies, they observed that nearly two-thirds of instruments had been developed and tested with college students, and many failed to provide detailed accounts of participants' gender, race, or age. These demographics are important because of systemic inequities linked to the underrepresentation of Black and Latinx youth and female youth pursuing CS in the United States (Margolis et al., 2017). Of the over 30 available standardized instruments included in their review, most had been adapted from fields outside of computing education (e.g., STEM) and thus required further validation to be applicable to computing. Additionally, many instruments did not capture the interrelated nature of different self-belief constructs (e.g., mindset, emotions, attitudes, values and others). As an exception, two popular instruments to measure students' self-beliefs—the *Instrument to Assess Student Self-Beliefs in CS* (IASS) (Scott & Ghinea, 2014) and *Computing Attitudes Survey* (CAS) (Dorn & Tew, 2015)—bring together several constructs related to student beliefs. However, these were both assembled and validated only for use with undergraduate students and provided little detail on participants' demographics. Thus, it is necessary to assemble and validate instruments to measure K-12 student beliefs that have age-relevant language and account for the specific characteristics of secondary school computing experiences (e.g., the use of physical computing, block-based programming, project-based learning, unplugged programming, and games; Waite & Sentance, 2021). Also missing from most instruments is an emphasis on creative personal expression through coding projects, at the core of much K-12 computing education for decades (Harel & Papert, 1991; Oleson et al., 2020). These aspects highlight the critical need for measuring secondary student self-beliefs in computer science learning contexts, particularly in areas of learning by designing.

In this paper we present a validated instrument called the *Computer Science Interest and Belief Inventory* (CSIBI) to assess secondary student self-beliefs in computing and their connection to anxiety, self-concept, and mindsets about computing. In the design of the CSIBI, we adopted a building talent approach (Margolis et al., 2016) to self-beliefs, that recognizes the key role of mindsets in organizing self-beliefs, to (1) identify five interrelated self-beliefs about CS (i.e., problem solving competency, fascination in design, value, creative expression, abilities), (2) investigate how CS self-beliefs and mindsets relate to anxiety and self-concept, and (3) use structural equation modeling to test a conceptual model where in mindsets function as mediators between CS self-beliefs and both programming anxiety and self-concept. To create the instrument, we followed best practice by adapting two existing validated instruments and creating a single new construct on CS creative expression to account for learning by designing. To wit, we adapted constructs in IASS (Scott & Ghinea, 2014) from undergraduate to secondary school students and from Java to e-textiles and Arduino and adapted items from the science-focused Activation Lab instruments (Dorph et al., 2016) to be CS-focused. The instrument includes nine total constructs on CS beliefs (i.e., problem solving competency beliefs, fascination in design, value of CS, CS creative expression, and e-textiles coding self-efficacy), programming mindsets (i.e., fixed and growth mindsets), and programming anxiety and programming self-concept. Our evaluation of the measurement instrument included survey responses of 303 students (grades 9-12) from four secondary schools with large percentages of historically marginalized students. In this paper we describe the development and validation of the instrument, the conceptual model with mindsets as mediators, and how this instrument could be used and/or adapted in other K-12 settings.

**Background**
Although computing education research has historically focused on cognitive aspects of learning, non-cognitive dimensions of learning such as motivation and emotion are instrumental in supporting all students (Lishinski & Yadav, 2019). Non-cognitive factors are not only interdependent with cognitive ones, but can also influence academic achievement, and general life outcomes (McGill et al., 2019). Measuring students' non-cognitive factors is important because pernicious belief systems, together with structural inequities, play a role in enabling or hindering the participation of students from historically marginalized backgrounds in computing (Lishinski & Yadav, 2019; Margolis et al., 2017). Within non-cognitive factors, a particular place has been assigned to self-beliefs. Self-beliefs are an array of different self-terms (Wylie, 1979)—including for instance, self-concept and self-efficacy—that share an emphasis on the beliefs individuals hold about their own abilities and attributes (Valentine et al., 2004). There is some evidence that self-beliefs influence academic achievement, with stronger effects when self-belief constructs are within the same domain as the measured academic area (Valentine et al., 2004). Indeed, research shows that self-beliefs are domain specific—i.e., self-beliefs about language arts do not necessarily apply to self-beliefs about mathematics—and multidimensional, including beliefs on cognitive, social and affective aspects of the self (Bandura, 1997; Harter, 1999). Dweck (2006) organized opposing self-beliefs into "fixed" and "growth" mindsets. Together with Yeager (2019) she argues that "some beliefs are not isolated ideas, but rather can serve an organizing function, bringing together goals, beliefs, and behaviors into what might be called a meaning system," (p. 483) proposing mindsets as the core of these meaning systems. Whereas learners with fixed mindsets perceive their intellectual abilities and social traits as unchangeable, those with growth mindsets believe their competencies and abilities can change and be developed over time. For instance, the assumption that some people are good at CS and others are not reflects a fixed mindset that often undermines the capacity of some students to persevere. On the other hand, learners with growth mindsets believe their competencies and abilities can change and be developed over time. Learners with growth mindsets tend to have higher resilience in challenging tasks, such as problem solving, than those with fixed mindsets (Yeager & Dweck, 2012). Indeed, mindsets influence how people choose challenges, persist over setbacks, and create value judgements (Dweck & Yeager, 2019). Mindsets are malleable and can change through interventions in which students learn about the potential to change their abilities and how to develop growth mindsets (Yeager & Dweck, 2012).

***Self-Beliefs and Mindsets in CS***
Student self-beliefs regarding computing play an important role in the complex process of learning computer programming. Within computing education research, self-beliefs related to efficacy, anxiety, enjoyment, confidence, belonging and persistence have been investigated (Decker & McGill, 2019). Some researchers such as Kinnunen and Simon (2010) propose that promoting a growth mindset may help decrease anxiety, hopelessness and attrition among novice programmers. Mindsets might be particularly helpful in persevering through problems and challenges during programming: while students with growth mindsets are likely to see bugs as opportunities for learning, those with fixed mindsets may become frustrated with failure and see it as a test of their intelligence (Burnette et al., 2020; Nolan & Bergin, 2016; Murphy & Thomas, 2008). Margolis and colleagues (Margolis et al., 2016) argue that addressing structural inequities and fostering growth mindsets are key to changing pernicious belief systems and broadening participation in computing. Whereas we acknowledge the importance of addressing structural inequities, in this paper we focus on students' self-beliefs. Margolis and colleagues (Margolis et al., 2016) recognize the important role of mindsets in mediating learner self-beliefs. This view is supported by Dweck's work in social psychology (Dweck, 2000; Dweck, 2006; Dweck & Yeager, 2019) that proposes that mindsets are at the core of belief systems and a growing number of studies on mindset in computing education that, with mixed results, investigate its relation to both specific cognitive and non-cognitive factors of learning (Apiola & Sutinen, 2020; Flannigan et al., 2022; Burnette et al., 2020; Quille & Bergin, 2020; Rangel et al., 2020; Woods, 2020; Gorson & O'Rourke, 2019; Stout & Blaney, 2017; Loksa et al., 2016; Lovell, 2014; Kench et al., 2016; Nolan & Bergin, 2016; Scott & Ghinea, 2013). Growth mindsets challenge traditional assumptions in CS about who can succeed in the discipline through the belief that "everyone can change and grow through application and experience" (Dweck, 2007, p.7). This is particularly important in disrupting notions that only some people are good at CS. Assuming a fixed mindset based on stereotypes or current representation in CS undermines the capacity of historically marginalized—particularly female, Black and Latinx—students to develop computational fluency and learn (Margolis et al., 2016). Indeed, as Apiola and Sutinen (2020) argue, mindset research in computing education should go beyond its relationship to academic achievement, addressing non-cognitive aspects of learning (such as self-beliefs) holistically through instruments validated in diverse cultural environments.

***Instruments to Study Students Self-Beliefs in CS***
While there is a growing interest in measuring computer science learner self-beliefs (Decker & McGill, 2019), few studies have investigated how different self-belief constructs relate to each other (Malmi et al., 2020). Instead, most attention has focused on individual constructs—such as mindset (e.g., Burnette et al., 2020, Rangel et al., 2020), self-efficacy (e.g., Steinhorst et al., 2020; Danielsiek et al., 2017) or anxiety (e.g., Nolan & Bergin, 2016). Few instruments bring together multiple, validated constructs related to student beliefs in the same instrument. Exceptions include the *Instrument to Assess Student Self-Beliefs in CS* (IASS) (Scott & Ghinea, 2014) and the *Computing Attitudes Survey* (CAS) (Dorn & Tew, 2015). IASS measured student self-beliefs about programming (aptitude mindset, anxiety, self-efficacy, self-concept, interest) building on control-value theory of achievement emotion (Scott & Ghinea, 2014). The idea is that learners' views of their ability (self-concept) and interest in computer science, alongside their beliefs about whether they can improve in computer science (i.e., growth versus fixed mindset) affect whether students become anxious while programming. This in turn affects the likelihood that learners will engage in programming. The instrument was validated with three cohorts of undergraduate students. CAS addressed discipline-specific attitudes with regards to problem solving in computer science (i.e., strategies, transfer, fixed mindset, confidence, and sense making), real world connections and enjoyment. CAS measures "novice to expert attitude shifts about the nature of knowledge and problem solving in computer science" (p. 1), such as the importance of understanding an algorithm while using it in programming. The instrument was validated and used in a pre-post format to test how it could measure attitudinal changes during an introductory undergraduate CS course (Dorn & Tew, 2015). The contributions of IASS and CAS are important, they were validated for use only with undergraduate students. This follows a trend where most measurement instruments available in CS education research are designed for college and university level students (Decker & McGill, 2019; Margulieux, 2020). Indeed, there is a need for comprehensive and holistic instruments designed with appropriate language and topics and validated to measure secondary school students' beliefs, as well as those of younger students.

### *Developing Instruments Specific to K-12 Students*
CS in K-12 schools has characteristics that must be considered in the design and validation of instruments. First, survey language needs to be age appropriate to ensure that learners understand the items presented. Too often survey instruments use complex terms related to self-beliefs or computing that may not be relevant or understandable for novice primary or secondary students. Second, construct items, particularly related to programming self-efficacy, need to be context appropriate. Programming learning is context-specific, especially for novices who may only have used one or two programming environments or languages. Thus, constructs for self-efficacy designed for programming languages or environments used in introductory undergraduate environments will not apply to many K-12 settings. For instance, the self-efficacy items in IASS (Scott & Ghinea, 2014) focus on Java, the language used in introductory CS courses at the universities they studied. However, many K-12 use other environments and languages, such as Scratch or other block-based coding environments, Python, Arduino, and others. Items on self-efficacy and/or ability with coding need to be based on students' learning experiences.

In addition, many K-12 CS settings focus on learning by designing, from the early days of Papert's work with turtles, Logo, and children to contemporary programming environments like Scratch (Resnick et al., 2009) or MakeCode (Ball et al., 2019) and standards and curricula (Caspersen, 2022; Waite & Sentence, 2021). In a recent review of K-12 approaches and strategies to CS education, Waite and Sentance (2021) highlight how the contexts and environments for learning programming in secondary school have been devised to increase learner interest and motivate learners to further pursue computing. As such many educators have created design environments where students pursue creativity, personal expression, and community relevance through programming (Oleson et al., 2020). Thus, it is common for secondary students to engage with physical computing using programmable microcontrollers (e.g., Przybylla & Romeike, 2014), transition from block-based programming environments to text-based programming languages (e.g., Weintrop & Wilensky, 2019), design personally meaningful projects in a variety of coding environments (e.g., Papert & Harel, 1991; Oleson et al., 2020) or program their own games (e.g., Kafai & Burke, 2015; Repenning et al., 2015). As such, a missing piece in many existing instruments on self-beliefs are items measuring students' creativity or personal expression with computing as well as fascination with design and self-efficacy in design environments.

Beyond the creation of more age-appropriate and context-relevant instruments for K-12 audiences, it is critical to validate all instruments—whether for undergraduates or K-12 students—with a broad range of demographics, especially with traditionally marginalized groups. Validating instruments with historically marginalized populations is crucial to conduct further research on how self-beliefs, in conjunction with systemic inequities, affect female, Black and Latinx students' participation in computing (Margolis et al., 2016). It is not adequate to use instruments with populations with whom the instruments have not been validated. This is particularly important regarding computer science, since one priority for developing instruments is to use them to evaluate interventions that aim to broaden participation in computing, i.e., with marginalized populations (Washington et al., 2016).

### *The Computer Science Interests and Beliefs Inventory (CSIBI) and Conceptual Model*

Adopting the perspective that mindsets are at the core of belief systems (Dweck, 2006; Dweck & Yeager, 2019; Margolis et al., 2016) we developed and validated an instrument for secondary school, novice students in a learning by designing intervention, examining self-beliefs in CS and their connections to programming mindset, programming anxiety, and self-concept.

#### *Instrument Constructs*

CSIBI is organized into three sets of constructs. First, the CS Beliefs contains 18 items across five constructs: problem solving competency beliefs, fascination in design, value of CS, CS creative expression, and e-textiles coding self-efficacy. Four of the five constructs were previously tested with 15 classrooms of secondary students in 14 schools (including the four schools with the four teachers in this study) (Kafai et al., 2019); the last construct was adapted to e-textiles coding self-efficacy for the present study (Scott & Ghinea, 2014). A second set of constructs includes seven items for mindsets (programming fixed mindset and programming growth mindset), and a third includes seven items for outcomes (programming anxiety and programming self-concept). All items were modified to be age appropriate as needed. Each construct demonstrated good reliabilities in their previous studies. Below we further explain each construct (see Table 1 for individual items).

Table 1. CSIBI constructs and their respective items.

| CSIBI Constructs and Items |
| --- |
| **Problem Solving Competency Beliefs** |
|  ● I think I am very good at: Figuring out how to fix things that don't work. |
|  ● I think I am very good at: Explaining my solutions to technical problems. |
|  ● I think I am very good at: Solving problems. |
|  ● I think I am very good at: Coming up with new ways to solve technical problems. |
|  ● I think I am very good at: Coming up with new ideas when working on projects. |
| **Fascination in Design** |
|  ● I love designing things! |
|  ● Designing new things makes me feel excited. |

- I talk about how things work with friends or family.

**Value of CS**
- Knowing computer science is important for contributing to my community.
- Knowing computer science is important for me in the future.
- Thinking like a computer scientist will help me do well in (none/a few/most/all of) my classes.
- I want to learn as much as possible about computer science.

**CS Creative Expression**
- I can be creative in computer science.
- I can express myself in computer science.
- I can make things that are interesting to me in computer science.

**E-textiles Coding Self-efficacy**
- I am confident that I can understand Arduino errors (e.g., was not declared in this scope, expected ';' before).
- I am confident I can write code for a simple e-textiles project.
- I am confident I can create a functional program for an e-textiles project that uses both sensors and actuators (e.g., LEDs, speakers).

**Programming Fixed Mindset**
- I have a fixed level of programming ability, and not much can be done to change it.
- I can learn new things about software development, but I cannot change my basic ability for programming.
- To be honest, I do not think I can really change my ability for programming.

**Programming Growth Mindset**
- No matter who you are, you can significantly change your programming ability.
- I can always substantially change my programming ability.
- No matter how much programming ability I have, I can always change it quite a bit.
- I can change even my basic programming ability considerably.

**Programming Anxiety**
- I often worry that it will be difficult for me to complete debugging exercises.
- I often get tense when I have to debug a program.
- I get nervous when trying to solve programming bugs.
- I feel helpless when trying to solve programming bugs.

**Programming Self-concept**
- I learn programming quickly.
- I have always believed that programming is one of my best subjects.
- In my programming classes, I can solve even the most challenging problems.

*Problem solving competency beliefs* address learners' confidence in their ability to solve CS-based problems. Competency beliefs are "people's judgments of their capabilities to organize and execute courses of action required to attain designated types of performance" (Bandura, 1986, p. 391). These beliefs are essential for learning scientific content knowledge and play an important role in predicting learning outcomes particularly for girls (Vincent-Ruz & Schunn, 2017). Furthermore, these reflect learners' goals which can shape whether they have entity (fixed) or incremental (growth) views of intelligence (Dweck & Legget, 1988). We adapted selected items from the Activation Lab STEM Competency Beliefs scale (Chen et al., 2017) to CS (e.g., "I think I am very good at explaining my solutions to math problems" to "I think I am very good at explaining my solutions to technical problems").

*Fascination in design* captures positive affect, interest, and curiosity towards design activities. Fascination refers to the emotional and cognitive attachment that learners have about topics which may intrinsically motivate participation and drive career interests (Dorph, et al., 2018). Items for this construct, notably about design (versus math or science), were drawn from the Activation Lab STEM Fascination scale (Chen et al., 2017). An example item is "Designing new things makes me feel excited."

*Value of CS* construct measures learners' beliefs about the value of CS. Learners can value CS by giving importance to how CS can help them achieve personal and societal goals. We adapted selected items from the Activation Lab STEM Values scale (Chen et al., 2017) to CS (e.g., "Thinking like a scientist will help me do well in all of my classes" to "Thinking like a computer scientist will help me do well in all of my classes"). Both value and fascination can influence student mindsets and beliefs about their capacity to learn and develop their CS abilities, intrinsically motivating learners (Dweck, 2000).

The *CS creative expression* construct was designed to measure learner beliefs about being able express themselves creatively with CS. This construct, not drawn from any prior instrument outside of our own work, is of particular importance to K-12 students as designing personally relevant artifacts is widely used to promote student engagement (Harel & Papert, 1991; Oleson et al., 2020). At the same time, learning by making or designing supports adolescents' identity formation by providing outlets for malleable, growth-oriented, ways of thinking and developing (Karwowski & Kaufman, 2017). An example item is "I can be creative in computer science."

The *E-textiles coding self-efficacy* construct is the only context-specific construct for the particular content in our intervention, namely using Arduino to code e-textiles. It captures learners' beliefs and assessments of their own ability to create and debug code in e-textiles projects. Self-efficacy has implications in how students perceive goals that can be reflected in their mindsets (Dweck & Legget, 1988). We adapted items from IASS (Scott & Ghinea, 2014) on debugging self-efficacy with Java in CS1 to e-textiles (e.g., "I am confident that I can understand Java exceptions (e.g., NullPointerException)" to "I am confident that I can understand Arduino errors (e.g., was not declared in this scope, expected ';' before)"). This context specific construct is based on Bandura's (1977) theoretical construct of personal efficacy. It can be easily adapted to more general physical computing or other coding contexts such as learning with Scratch or Python.

The *programming fixed mindset* construct includes items related to learner beliefs of programming ability being fixed and not malleable. We reviewed several constructs from existing CS mindset instruments (Scott & Ghinea, 2014; Burnette et al, 2020) and Dweck's (2000) general mindset measures. We decided to adapt IASS's construct (Scott & Ghinea, 2014), which replaced "intelligence" in items from Dweck's (2000) instrument for "programming aptitude", and edited it following Burnette and colleague's (2020) proposal of centering ability (e.g., "To be honest, I do not think I can really change my aptitude for programming" to "To be honest, I do not think I can really change my ability for programming") to use language more familiar to students.

The *programming growth mindset* construct includes items related to learner beliefs of programming ability being able to change and be developed over time. We adapted items from Dweck's (2000) general mindset instrument replacing "intelligence" for "programming ability" (e.g., "No matter how much intelligence you have, you can always change it quite a bit" to "No matter how much programming ability I have, I can always change it quite a bit").

The *programming anxiety* construct captures learner's anxiety towards programming and debugging. Learning to code, and encountering bugs, can generate emotional responses related to anxiety such as fear that can lead to disengagement and the avoidance of computer programming (Nolan & Bergin, 2016). Mindset can influence learner anxiety, where incremental views of intelligence relate to lower anxiety (Smith & Capuzzi, 2019). The items in the programming anxiety construct were drawn from IASS (Scott & Ghinea, 2014) which in turn adapted Wiegfield and Meece's (1988) math anxiety construct to CS: e.g., I often worry that it will be difficult for me to complete debugging exercises.

The *programming self-concept* construct includes items that represent beliefs learners may have about their ability to be good programmers. This construct highlights the affective or emotional elements of being a programmer versus a more "cognitive assessment of success" with programming (see Scott & Ghinea, 2014, p.125 for an explanation of the differences between self-concept and self-efficacy in programming). Ability-based self-concept and competency-based achievement goals seem to vary depending on learners' mindsets (Heyder et al., 2020; Dweck 1988; Dweck 2000). The items for this construct were drawn from IASS (Scott & Ghinea, 2014) which adapted Ferla and colleagues' (2009) and Eccles and Wigfield's (1995) self-concept items to CS: e.g., In my programming classes, I can solve even the most challenging problems.

*Proposed Model*

We propose and investigate the stability of the CSIBI, and test whether and how the five CS Beliefs (problem solving competency, fascination in design, value of CS, CS creative expression, and e-textiles coding self-efficacy) influence programming mindsets regarding innate versus learned programming ability, hypothesizing that mindsets function as mediators that directly influence programming anxiety and programming self-concept (see Figure 1.). Our hypothesis builds on previous work outside of CS education (described for each of the constructs in the previous section) that suggests that competency beliefs, fascination, value, creative expression, and self-efficacy can shape learner goals and influence their mindsets (Dweck & Legget, 1988; Dweck, 2000; Karwowski & Kaufman, 2017) and research on the relationships between mindsets, self-concept and anxiety (Smith & Capuzzi; 2019, Heyder et al., 2020; Dweck 1988; Dweck 2000). In this paper we present the instrument validated with secondary students completing a learning by designing CS unit and test the hypothesized model.

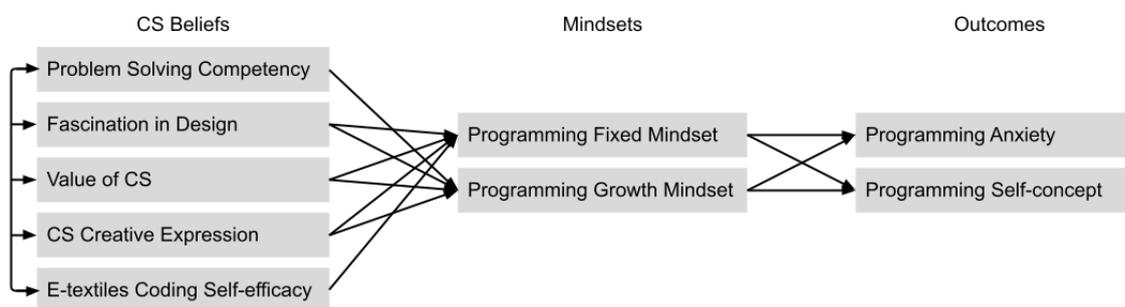

*Figure 1*. CSIBI hypothesized model, with mindsets as mediators, anxiety, and self-concept as outcomes.

**Methods**

*Context and Participants*

Data used to validate the CSIBI was collected from 11 classrooms in four secondary school schools in two metropolitan school districts on the West Coast of the United States during Spring 2021. Among the 303 secondary school students who completed the survey, 30.8% self-identified as girls, 56% self-identified as boys, 0.6% as non-binary, and 12.4% did not provide gender information. Most students (82%) spoke a language other than English at home, 55.9% had no previous computer science experience, and 73.7% had family members with at least some college education. School level demographics (See Table 2) show that at least 75% of the targeted student population were non-White and at least 49% were eligible for free/reduced cost lunch. Of note, we intentionally did not collect race or ethnicity demographic information in this survey because of a decision to waive consent as part of the data collection process. This was a conscious choice to support broader participation in the survey at a time when schools in the study were entirely virtual due to health guidelines regarding the COVID-19 pandemic. In discussion with Institutional Review Board authorities, the decision was made to note students' teacher and class period as well as student gender; however together we felt that requesting racial/ethnic demographic information could make data identifiable to a single student (e.g., it would not be difficult to identity the one male Filipino student in a particular class).

Table 2. Demographics of schools where participating Exploring Computer Science classes took the survey.

| School ID | Grade | Number of Students | Gender | Ethnicity | | English Learners | Reduced Lunch/ Socioeconomically Disadvantaged |
|---|---|---|---|---|---|---|---|
| School 1 (2 classes) | 9-12 | 4739 | F-49% M-51% | AA-3.6%, As-17.5%, HL-39.8%, Wh-25.5%, NR-2% | AI-0.36%, Fi-9.7%, PI-0.3%, 2+-1.2%, | 3% | 49% |
| School 2 (1 class) | 9-12 | 1573 | F-64% M-36% | AA-42%, As-0.6%, HL-55.7%, Wh-0.9%, NR-0.13% | AI-0.25%, Fi-0.06%, PI-0.06%, 2+-0.2%, | 2% | 91% |
| School 3 (5 classes) | 9-12 | 3196 | F-51% M-49% | AA-19.4%, As-8.4%, HL-45.8%, Wh-15%, NR-1.6% | AI-0.3%, Fi-4.8%, PI-2.8%, 2+-1.9%, | 5% | 58% |
| School 4 (3 classes) | 9-12 | 1263 | F-45% M-55% | AA-0.16, As-0.4%, HL-98.4%, Wh-0.71% | AI-0.08, Fi-0.16, PI-0.08, | 19% | 95% |

NOTE: F=Female; M=Male; AA=African American; AI=American Indian or Alaska Native; As=Asian; Fi=Filipino; HL=Hispanic/Latino; PI=Pacific Islander; Wh=White; 2+=two or more races; NR=not reported.

All participants were enrolled in Exploring Computer Science (ECS), a year-long, equity-focused, inquiry-based introductory computing course for secondary school students that has been adopted by school districts across the United States (Goode et al., 2012). It includes six curricular units on (1) Human-Computer Interaction; (2) Problem Solving; (3) Web Design; (4) Introduction to Programming; (5) Robotics or e-textiles; (6) Computing Applications (Goode & Margolis, 2011). The ECS program also includes teacher and multi-stake holder professional development, including school counselors and administrators, centered on broadening participation in computing (Goode et al., 2012; Flapan et al., 2020). This is particularly important to develop teachers' pedagogical content knowledge, culturally relevant practices and equitable access to computing. Prior research has found that ECS teachers engage in practices that broaden participation by demystifying computing, showing its connections to everyday life, addressing social issues in CS and students' communities, and valuing student voices (Ryoo, 2019). At the same time professional development for counselors is particularly important as they often play the role of gatekeepers by encouraging or discouraging students to take computing classes based on implicit biases, by working with counselors and teachers ECS aims to invite all students, particularly those from historically marginalized backgrounds to enroll in the course (Flapan et al., 2020; Margolis, 2017). Within ECS, the development of growth mindsets among students, instructors, and counselors intentionally plays an essential role, since it aims to prioritize equity by recognizing that all students who have access to high-quality education can improve their ability and participate in computing (Margolis et al., 2017). For ECS student success is not only measured by knowledge gained, but also by student attitudes and beliefs about computing in their lives; this involves attending to pernicious belief systems that may limit participation of historically marginalized students in computing (Margolis et al., 2017).

In this study, students completed the survey toward the end of the school year, at the beginning of the ECS 12-week long e-textiles unit (http://exploringcs.org/e-textiles), designed for youth to create personally relevant creative projects while learning new coding, circuitry, and crafting technical skills (see Kafai et al., 2019).

*Data collection*
CSIBI was administered as an online survey (via Qualtrics) with items and constructs presented in randomized sequences, and demographic questions coming at the end (see Appendix 1). For each of the nine-constructs, responses from participants were recorded using a four-point Likert scale to encourage greater reflection and avoid neutral or indifferent responses, ranging from strongly disagree (1) to strongly agree (4). There was one exception, the item "Thinking like a computer scientist will help me do well in": here scale labels went from none of my classes (1) to all of my classes (4). The CSIBI also captured demographic information on gender identity, previous experience with computing, language spoken at home and family college attendance history (see Table 3). Teachers created assignments that required students to complete our online survey in their school's learning management systems (LMS). Students were given time to complete the survey during class and also allowed to do it as homework within a certain time period. The instrument was administered as a pre-test at the beginning of the e-textiles unit with the intention of administering a post-test at the end of the unit. (Of note, due to the constraints of the study conducted virtually during the COVID-19 pandemic, we did not collect identifiers that would have enabled pre/post connections, and there was significant attrition in participation at the end of the school year; an analysis of any connecting surveys across time and different sample sizes, pre/post comparisons are beyond the scope of this study[1]). In this paper we introduce the instrument and model with pre-test data to establish the validity of using it in studies.

Table 3. Demographic information collected in CSIBI.

| Construct | Item |
|---|---|
| Previous experience with computing | Before this computer science class, did you take any computer science classes or workshops? Yes/No |
| Gender | Please indicate your gender: Female/ Male/ Other/ Decline to indicate |
| Home Language | How often do people in your home talk to each other in a language other than English? Never/ Once in a while/ About half of the time / Most of the time/ All the time |
| Family College Attendance History | Who in your immediate family has attended college? (Select all that apply): Mother / Father / Sibling / Grandparent / Other (please specify) / No one |

*Analysis*

To assess the nine-factor structure for the 32-item Computer Science Interests and Beliefs inventory (CSIBI), a maximum likelihood confirmatory factor analysis (CFA) was conducted (using Mplus 8). The CFA analysis tests whether the number of proposed factors fits the data, each indicator item adequately loads onto the proposed factor, the errors (or uniqueness) across each indicator are unrelated, and the relationship (correlations) exist across all the latent factors.

Following this analysis, structural equation modeling (SEM) was used to test how the hypothesized model connected CS Belief constructs to growth and fixed mindsets, and the mindsets' mediating role in predicting programming anxiety and programming self-concepts. As with CFA, SEM examines the factor structure of each latent variable (i.e., whether and how each indicator adequately loads onto the factor). In addition, SEM tests the proposed connections and directionality among a series of indicator items and latent variables and provides a framework to identify mediating relationships between variables.

In general, model indices are used for CFA and SEM to compare differences between the hypothesized model (specific relationship among variables) and models that are better or worse (where the relationships among variables differ or do not exist as hypothesized). As there are no definitive model fit criteria for CFA, multiple indicators were used. For this study, adequate model fit indicators were chi-square goodness of fit (at $p > .05$ or $x^2/df < 3.00$), root mean square error of approximation (RMSEA < .06), standardized root mean square residual (SRMR < .08), comparative fit index (CFI > .90) and Tucker–Lewis index (TLI > .90).

---

[1] Because of the timing of the study during Spring 2021 with schools still operating virtually due to the COVID-19 pandemic, there was significant attrition for the post-tests. For a study with limited findings studying only the post-test across an intervention and comparison groups, see Morales-Navarro et al., 2023.

## Findings

To test the CSIBI constructs and the conceptual model, we conducted two separate analyses: confirmatory factor analysis (CFA) and structural equation modeling (SEM). While SEM tests both the measurement stability of the factors with their items and proposed directional and mediational influences of the proposed model, a separate CFA was chosen to highlight and validate the CS Beliefs constructs.

### CS Interest and Beliefs Inventory

Supporting the development and validation of the CSIBI, confirmatory factor analysis results indicated that the nine-factor model was a good fit for the data across model indices (except the $x^2$ p-value, which is often significant with larger sample sizes as is the case in this study; see model indicators in Table 4). In addition, the factor loadings for each item averaged .742 (ranging between .491 to .875; with two loadings below .600). All loadings were significant ($p < .001$). See Table 4 for factor loadings and reliabilities for all constructs.

Table 4. Factor loadings and reliabilities across all variables (N = 303).

| | Factor Loadings | S.E. | Factor Determinacy | Cronbach's α |
|---|---|---|---|---|
| **Problem Solving Competency Beliefs** | | | | |
| I think I am very good at: Figuring out how to fix things that don't work. | 0.735 | 0.032 | 0.933 | 0.836 |
| I think I am very good at: Explaining my solutions to technical problems. | 0.678 | 0.036 | | |
| I think I am very good at: Solving problems. | 0.715 | 0.033 | | |
| I think I am very good at: Coming up with new ways to solve technical problems. | 0.797 | 0.027 | | |
| I think I am very good at: Coming up with new ideas when working on projects. | 0.642 | 0.039 | | |
| **Fascination in Design** | | | | |
| I love designing things! | 0.759 | 0.032 | 0.925 | 0.720 |
| Designing new things makes me feel excited. | 0.876 | 0.028 | | |
| I talk about how things work with friends or family. | 0.491 | 0.05 | | |
| **Value of CS** | | | | |
| Knowing computer science is important for contributing to my community. | 0.684 | 0.038 | 0.918 | 0.778 |
| Knowing computer science is important for me in the future. | 0.717 | 0.036 | | |
| Thinking like a computer scientist will help me do well in (none/a few/most/all of) my classes | 0.624 | 0.042 | | |
| I want to learn as much as possible about computer science. | 0.705 | 0.037 | | |
| **CS Creative Expression** | | | | |
| I can be creative in computer science. | 0.816 | 0.024 | 0.951 | 0.870 |
| I can express myself in computer science. | 0.819 | 0.023 | | |
| I can make things that are interesting to me in computer science. | 0.867 | 0.02 | | |
| **E-textiles Coding Self-efficacy** | | | | |
| I am confident that I can understand Arduino errors (e.g., was not declared in this scope, expected ';' before). | 0.764 | 0.031 | 0.924 | 0.814 |
| I am confident I can write code for a simple e-textiles project. | 0.756 | 0.032 | | |
| I am confident I can create a functional program for an e-textiles project that uses both sensors and actuators (e.g., LEDs, speakers). | 0.791 | 0.029 | | |
| **Programming Fixed Mindset** | | | | |
| I have a fixed level of programming ability, and not much can be done to change it. | 0.701 | 0.049 | 0.897 | 0.749 |
| I can learn new things about software development, but I cannot change my basic ability for programming. | 0.581 | 0.051 | | |
| To be honest, I do not think I can really change my ability for programming. | 0.833 | 0.049 | | |
| **Programming Growth Mindset** | | | | |
| No matter who you are, you can significantly change your programming ability. | 0.733 | 0.031 | 0.945 | 0.870 |
| I can always substantially change my programming ability. | 0.827 | 0.023 | | |
| No matter how much programming ability I have, I can always change it quite a bit. | 0.759 | 0.029 | | |
| I can change even my basic programming ability considerably. | 0.813 | 0.024 | | |
| **Programming Anxiety** | | | | |
| I often worry that it will be difficult for me to complete debugging exercises. | 0.791 | 0.029 | 0.931 | 0.786 |
| I often get tense when I have to debug a program. | 0.748 | 0.031 | | |
| I get nervous when trying to solve programming bugs. | 0.842 | 0.025 | | |
| I feel helpless when trying to solve programming bugs. | 0.653 | 0.038 | | |
| **Programming Self-concept** | | | | |
| I learn programming quickly. | 0.761 | 0.034 | 0.910 | 0.789 |
| I have always believed that programming is one of my best subjects. | 0.704 | 0.037 | | |
| In my programming classes, I can solve even the most challenging problems. | 0.776 | 0.033 | | |

*Note*: All factor loadings were significant (p < .001)

Examining the relationship among the nine-constructs at the bivariate correlation level (see Table 5), five CS Beliefs constructs (problem solving competency, fascination, value, CS creative expression, e-textiles coding self-efficacy) were found to be significantly positively related to each other. For instance, as expected, increased competency beliefs were significantly correlated with increases in fascination in STEM, value of CS, CS creative expression, and e-textiles coding self-efficacy. In addition, these five CS Beliefs constructs (competency, fascination, value, CS creative expression, e-textiles coding self-efficacy) were found to significantly relate to higher growth mindset; however, none of the CS Beliefs constructs correlated with fixed mindset. In addition, the five CS Beliefs significantly correlated with increased programming self-concept but were not associated with programming anxiety.

Table 5. Correlation (standard error) table among all factors (N = 303).

| | 1 | 2 | 3 | 4 | 5 | 6 | 7 | 8 | 9 |
|---|---|---|---|---|---|---|---|---|---|
| 1. Problem Solving Competency Beliefs | -- -- | | | | | | | | |
| 2. Fascination with Design | 0.661*** (.046) | -- -- | | | | | | | |
| 3. Value of CS | 0.583*** (.053) | 0.623*** (.052) | -- -- | | | | | | |
| 4. CS Creative Expression | 0.603*** (.047) | 0.597*** (.049) | 0.787*** (.037) | -- -- | | | | | |
| 5. E-textile Coding Self-efficacy | 0.607*** (.049) | 0.486*** (.058) | 0.685*** (.047) | 0.746*** (.038) | -- -- | | | | |
| 6. Programming Fixed Mindset | -0.009 (.078) | -0.086 (.074) | -0.083 (.077) | -0.052 (.077) | 0.115 (.081) | -- -- | | | |
| 7. Programming Growth Mindset | 0.618*** (.046) | 0.490*** (.055) | 0.646*** (.047) | 0.739*** (.036) | 0.582*** (.050) | -0.041 (.061) | -- -- | | |
| 8. Programming Anxiety | -0.044 (.048) | -0.068 (.042) | -0.077 (.049) | -0.07 (.053) | 0.009 (.047) | 0.416*** (.065) | -0.083 (.065) | -- -- | |
| 9. Programming self-concept | 0.396+ (.047) | 0.300*** (.047) | 0.401*** (.048) | 0.466*** (.047) | 0.394*** (.047) | 0.146 (.080) | 0.636*** (.048) | -0.326*** (.064) | -- -- |
| Means | 2.854 | 2.930 | 2.797 | 2.950 | 2.493 | 2.411 | 2.963 | 2.645 | 2.336 |
| SD | 0.521 | 0.569 | 0.556 | 0.652 | 0.643 | 0.625 | 0.557 | 0.609 | 0.648 |

+p <.10, ***p <.001

### *Conceptual Model*

Structural equation modeling (SEM) results supported the factor structure of the 9 constructs and partially supported the hypothesized model, demonstrating the importance of mindsets in predicting anxiety and self-concept. At the general level, model fit indices showed that the hypothesized model fit the data well (see model indices in Table 6). However, path analysis results revealed that three of the five CS Belief constructs differentially predicted mindsets. That is, only problem-solving competency and CS creative expression predicted significant increases in growth mindset among the five CS Belief constructs, and although bivariate results show that CS factors were not related to fixed mindset, beliefs about e-textiles coding self-efficacy emerged as a significant predictor of increased fixed mindset. This later finding is likely related to the interrelationship among the five constructs. Lastly, growth mindset was also found to significantly encourage programming self-concept, while fixed mindset served a dual role in increasing programming anxiety and increasing programming self-concept.

Table 6. Fit Indices, criteria, and results for CFA and SEM (N = 303).

| Fit Index | Adequate Fit Criteria | CFA Results | Structural Equal Model Results |
|---|---|---|---|
| X2 (df =125) | N/A | 697.24 (df = 428) | 852.090 (df = 439) |
| X2 / df | < 3.00 | 1.629 | 1.941 |
| p | > .05 | < .001 | < .001 |
| TLI | > .90 | .934 | .901 |
| CFI | > .90 | .943 | .913 |
| SRMR | < .08 | .057 | .086 |
| RMSEA | < .06 | .046 | .056 |
| RMSEA 90% CI | | .039, .052 (p = .884) | .050, .061 (p = .046) |

To determine whether growth and fixed mindsets function as mediators between the CS Belief constructs and both programming anxiety and self-concept, indirect effects were calculated. Results show that programming growth mindset significantly predicting programming self-concept and functioned as a significant mediator between problem solving competency beliefs and programming self-concept [indirect effect B = .198, 95% CI (.089, .308), SE = .056, p < .001] and between CS creative expression and programming self-concept [indirect effect B = .353, 95% CI (.203, .504), SE = .077, p < .001]. Programming fixed mindset was only a mediator between e-textiles coding self-efficacy and programming anxiety [indirect effect B = .162, 95% CI (.046, .279), SE = .060, p = .006], but not between e-textiles coding self-efficacy and programming self-concept [indirect effect B = .068, 95% CI (-.004, .140), SE = .037, p = .066]. All other indirect effects were non-significant. The full model is displayed in Figure 2.

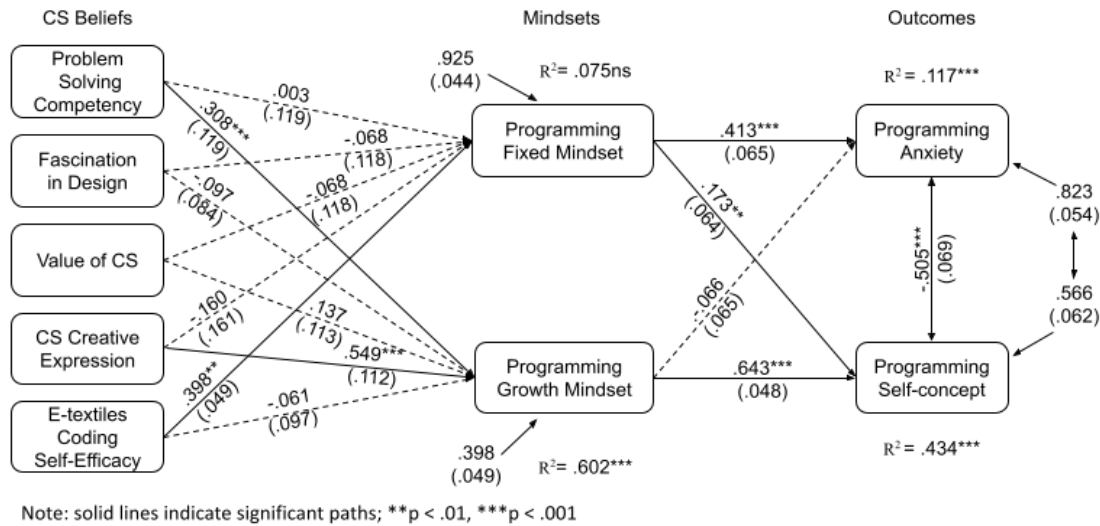

*Figure 2.* Structural equation modeling results (standardized betas and SE).

In sum, our conceptual model connecting CS Beliefs, mindsets, programming self-concept and anxiety was partially supported. Consistent with Margolis and colleagues' (Margolis et al., 2016) argument, growth mindsets about programming ability connected with CS beliefs constructs and emerged as a predictor and promoter of programming self-concept. In addition, growth mindset functions as a mediator explaining the relationship between some CS belief factors (i.e., problem solving competency and CS creative expression). In contrast, fixed mindsets beliefs about programming ability generally did not relate to CS Beliefs; however, they did predict programming anxiety and self-concept.

**Discussion**

***Towards a Validated Instrument for Novice Secondary Students***
In this paper we presented and validated an instrument to measure secondary school students' self-beliefs in CS. The *Computer Science Interests and Beliefs Instrument* (CSIBI) includes nine validated constructs: CS beliefs (five constructs including problem solving competency beliefs, fascination in design, value of CS, CS creative expression and e-textiles coding self-efficacy), programming mindsets (two constructs including fixed mindset and growth mindset), and emotional and self-concept outputs related to programming more broadly (two constructs including programming anxiety and programming self-concept). Of note, the five constructs about CS values and beliefs were significantly correlated with the programming growth mindset and not the fixed mindset, providing evidence for the crucial role of mindset in organizing self-beliefs. Together, these constructs provide a well-rounded instrument that can be used to assess secondary school students' self-beliefs about computing and programming, especially in contexts of learning by designing.

CSIBI provides a validated instrument that fills several gaps in CS measurement tools by focusing on secondary students in contexts of learning by designing. As Margulieux and colleagues (2020) noted, existing validated inventories or surveys of self-beliefs have largely focused on undergraduate students (Scott & Guinea, 2014; Tew & Dorn, 2012) and generally provide few demographics relevant to historically marginalized populations in CS. CSIBI responds to the dramatic expansion of computing education in K-12, focusing on secondary students, specifically high school students (grades 9-12), designing creative computing artifacts. To create CSIBI, we built on constructs from existing instruments and adapted them in specific ways. First, we made the language more accessible, for instance changing "aptitude" to "ability"—a term more accessible to secondary students. Second, we included a construct on fascination with design and further developed a new construct on CS creative expression, addressing a common feature of secondary school CS experiences that aim to promote learning CS through the design of personally relevant artifacts (Oleson et al., 2020). Such items, while possibly relevant to any student, are essential in contexts of learning by making or designing and to support adolescents' identity formation. Third, we developed context-specific self-efficacy questions for a particular application of computing, namely physical computing, to assess students' sense of ability with the e-textiles curriculum. The specificity of the e-textiles coding self-efficacy construct complements other constructs about students' consideration of programming or computing more broadly. These changes were made because nearly all students in our study—like many K-12 students—were new to computing, as well as to respond to the common goal of K-12 CS education to increase learner interest and motivate learners to further pursue computing. Further, we validated our instrument with novice secondary students from several schools, all of which had majority populations of historically marginalized students. The demographic questions in our instrument—about gender, family education, and English language use at home—provide critical information missing in prior studies. Requesting this information and validating instruments with marginalized populations is particularly important given the driving objectives in much of CS education to broaden participation in the field.

Designed to measure student self-beliefs in the context of ECS or similar CS learning environments that prioritize student interest, beliefs, and mindsets alongside content learning, the instrument attends to attitudes and beliefs that students have about themselves in relation to CS. The instrument is of great relevance to ECS since one of the goals of the program is to attend to the pernicious belief systems that may limit participation of historically marginalized students in computing (Margolis et al., 2017). Within this context measures of beliefs, mindsets and their outcomes are as important as and complementary to traditional measures of student achievement.

The instrument can be broadly applicable to other secondary education contexts where computing focuses on learning by designing. For instance, its focus on design and creative expression aligns well with the first Big Idea of CS of the AP CS Principles which centers on creativity (Astrachan & Briggs, 2012) as well as Code.org's CS Discoveries activities that predominantly involve design (Oleson et al., 2020). This is important because while creativity and design are often encouraged in K-12 computing education to increase student engagement (Waite & Sentance, 2021) these are understudied constructs (Oleson et al., 2020). CSIBI provides standardized measures that can be used to further investigate student's beliefs about design, creativity and computing across curricula.

It is worth noting that only one construct, e-textiles coding self-efficacy, is specific to a particular programming language and computing context. Just as we adapted items in IASS (Scott & Ghinea, 2014) from Java to Arduino and e-textiles to create the items in this construct, others could adapt the items to be relevant to Scratch, Python, MakeCode, or another programming environment and validate the construct specific to their contexts. The remaining constructs are broadly applicable to many interventions and computing contexts. Similarly, validating the instrument with students in programs other than ECS could be helpful to then be able to compare results across curricula. We look forward to future applications and adaptations of CSIBI (including by age group) and believe the instrument will support the rigor and validity of research on students' engagement with computing.

*A Validated Instrument for Novice Secondary Students*
Our work in developing the CSIBI for K-12 students responded to the call issued by Malmi and colleagues (Malmi et al., 2020) for more validated instruments to increase our holistic understanding of learner beliefs, emotions, and attitudes. Structural equation modeling results demonstrated our hypothesized model in which mindsets mediated other self-beliefs about computing. In the model, problem solving competency beliefs and CS creative expression promoted programming growth mindset, which subsequently fostered students' programming self-concept. E-textiles coding self-efficacy connected to a greater fixed mindset, which in turn predicted programming anxiety and programming self-concept. While the connection between e-textiles coding self-efficacy and programming fixed mindset was unexpected, we hope to further investigate it in future applications of the instrument. The results and hypothesized relationship between constructs should be pursued in further studies to better understand how programming mindsets relate to other self-beliefs (see also next section). Here further research is necessary to confirm the relationships between constructs and investigate how the instrument captures changes over time or in learning contexts. As Margulieux and colleagues (2019) argue, for CS education research it is crucial to develop and standardize domain-specific instruments to allow for comparisons across studies and work towards a strong evidence-based research agenda. In the following section, we discuss strategies for developing validated instruments for K-12 demographics and suggest areas which require further research and development to address this crucial need.

By presenting a model of how these constructs relate to each other, we hope to contribute to a holistic understanding of self-beliefs (Malmi et al., 2020) that builds on Margolis and colleagues' (Margolis et al., 2016) proposal that mindsets mediate student CS belief systems. We hope that CSIBI can be used to measure and evaluate interventions that aim to decrease attrition and broaden participation in CS. At the same time, we recognize that solely promoting students' growth mindsets is not enough, to broaden participation in computing it is necessary to investigate and address the structural inequalities that prevent students from accessing and pursuing computing education.

*Further Research in Instrument Development*
The validated CSIBI instrument captures many salient self-beliefs relevant to learning CS for secondary school students, especially in the context of electronic textile design. We invite researchers to examine and adapt CSIBI for other computing contexts. Because only one construct is specific to e-textiles, CSIBI should already provide a means for comparisons across studies in secondary school contexts. We believe the wording of questions is age-appropriate for grades 6-12, but having only validated it with grades 9-12, its use with younger audiences should be considered carefully. We look forward to seeing how CSIBI can be adapted, applied to, and compared across other CS education contexts.

In addition, future iterations of the CSIBI and similar instruments should seek to address under-researched topics such as learner beliefs about collaboration and ethical issues in computing. While self-belief surveys conceptually (and by design) focus on individual students, learning in secondary school classrooms is often collaborative. Margulieux and colleagues (2019) have called explicitly for more research assessing collaboration as part of CS learning. In fact, research studies of electronic textiles in secondary school classrooms find collaboration to be a key aspect in how students learn (Fields et al., 2018). Instruments such as CSIBI should also address student beliefs on collaboration, for example, do students believe collaborating with peers will help them improve their projects or do they consider CS to be a collaborative space (Shaw et al., 2019).

Related, future developments of self-belief instruments should include beliefs on ethical or critical issues around computing. While many assessments focus on learning computational concepts and practices, there is now a growing recognition of the importance of addressing the good and the harm computing technologies can inflict, as well as the lack of diversity in the field at large (Kafai & Proctor, 2022). As such it is crucial to investigate student self-beliefs about these issues. One possible direction could be to examine how students connect learning computing with social implications and how this connects to their interests. Do students believe it is important to reflect on the implications and limitations of computing and of the projects they create?

Finally, further research should consider the CSIBI (or similar) instruments in relation to practice. Student practices do not always align with self-reported instrument responses (Gorson & O'Rourke, 2019). Indeed, a growth mindset may not even be consistently beneficial for students if structural inequities are not addressed (Margolis, 2016; Stout & Blaney, 2017). Studies could take a multi-pronged approach to compare student mindsets as measured in instruments like CSIBI and compare them to growth mindset practices (Campbell et al., 2020; Morales-Navarro et al., 2021) to better understand connections, gaps, and tensions between self-report and actual growth mindset practices such as persisting after setbacks and choosing challenges to learn more.

**Conclusions**

In this paper we presented CSIBI, a validated instrument to measure secondary school students' self-beliefs in CS. The instrument developed provides a critical contribution to the field of K-12 computing education research presenting a tool to holistically measure non-cognitive factors of learning computing in creative, project based-learning contexts. We hope others can adapt and develop inventories that expand it into other contexts of K-12 CS education in order to investigate self-beliefs of different student populations. This is particularly needed for historically marginalized students to better understand how they connect to computing and where there are gaps or tensions in students' development self-beliefs about computing and how these tensions relate to systemic inequities. We found that problem solving competency beliefs, fascination in design, value of CS, CS creative expression and e-textiles coding self-efficacy were significantly correlated to programming growth mindset—and not fixed mindset—providing evidence for the role of mindset in organizing self-beliefs. Our analysis provides an initial model for how constructs of self-beliefs are organized, with mindsets as mediators of broader programming anxiety and self-concept. This points to the continued importance of work and investigation into self-beliefs and mindsets and their role in supporting (or preventing) learning in K-12 CS education. We suggest strategies and further research directions into adapting instruments to study of K-12 students' self-beliefs.

**Acknowledgments**
This work was supported by a grant from the National Science Foundation to Yasmin Kafai (#1742140). Any opinions, findings, and conclusions or recommendations expressed in this paper are those of the authors and do not necessarily reflect the views of NSF, Cal State University Pomona, the University of Pennsylvania, or Utah State University. Special thanks to Justice Toshiba Walker and Debora Lui for support in designing the survey instrument and to Gayithri Jayathirtha for her assistance in data collection.

**Research Ethics and Consent Statement**
We recruited students already enrolled in introductory computer science high school classes. A researcher visited the classes virtually to invite students to participate in the study. In discussion with Institutional Review Board authorities, we decided to waive consent as part of the data collection process. This was a conscious choice to support broader participation in the survey at a time when schools in the study were entirely virtual due to health guidelines regarding the COVID-19 pandemic. Since consent was waived, we did not collect individual identifiers and/or any racial/ethnic demographic information could make data identifiable to a single student. The school-wide demographic data included in this study was publicly available information. Students did not receive any incentives for participating in the study. Research protocols and data collection methods were approved by the IRB board of the University of Pennsylvania (Protocol: 827747).

**Data Availability Statement**
Due to the nature of this research, participants of this study did not agree for their data to be shared publicly. Supporting data is not available.

**Disclosure Statement**
No potential conflict of interest was reported by the authors.

**Funding details**
This work was supported by the National Science Foundation [1742140].

**References**
Apiola, M., & Sutinen, E. (2020, November). Mindset and study performance: New scales and research directions. In Proceedings of the 20th Koli Calling International Conference on Computing Education Research (pp. 1–9).
Astrachan, O., & Briggs, A. (2012, June). The CS principles project. ACM Inroads, 3(2), 38–42. https:/doi.org/10.1145/2189835.2189849
Ball, T., Chatra, A., de Halleux, P., Hodges, S., Moskal, M., & Russell, J. (2019, October). Microsoft makecode: Embedded programming for education, in blocks and typescript. In Proceedings of the 2019 ACM SIGPLAN Symposium on SPLASH-E, Athens, Greece (pp. 7–12).

Bandura, A. (1977). Self-efficacy: Toward a unifying theory of behavioral change. Psychological Review, 84(2), 191. https://doi.org/10.1037/0033-295X.84.2.191

Bandura, A. (1986). Social foundations of thought and action: A social cognitive theory. Prentice Hall.

Bandura, A. (1997). Self-efficacy: The exercise of control. WH Freeman, Times Books, Henry Holt & Co.

Burnette, J. L., Hoyt, C. L., Russell, V. M., Lawson, B., Dweck, C. S., & Finkel, E. (2020). A growth mindset intervention improves interest but not academic performance in the field of computer science. Social Psychological and Personality Science, 11(1), 107–116. https://doi.org/10.1177/1948550619841631

Campbell, A., Craig, T., & Collier-Reed, B. (2020). A framework for using learning theories to inform 'growth mindset' activities. International Journal of Mathematical Education in Science and Technology, 51(1), 26–43. https://doi.org/10.1080/0020739X.2018.1562118

Caspersen, M. E. (2022). Informatics as a Fundamental Discipline in General Education: The Danish Perspective. In H.Werthner, E.Prem, E. A. Lee, & C. Ghezzi (Eds.), Perspectives on Digital Humanism (pp. 191–200). Springer. https://doi.org/10.1007/978-3-030-86144-5_26

Chen, Y.-F., Cannady, M. A., Schunn, C., & Dorph, R. (2017a). Measures Technical Brief: Competency Beliefs in STEM. Activation Lab. http://www.activationlab.org/wp-content/uploads/2017/06/CompetencyBeliefs_STEMReport_20170403.pdf .

Chen, Y. F., Cannady, M. A., Schunn, C., & Dorph, R. (2017b). Measures technical brief: Values in STEM. Activation Lab.

Chen, Y.-F., Cannady, M. A., Schunn, C., & Dorph, R. (2017c). Measures Technical Brief: Values in STEM. Activation Lab. http://www.activationlab.org/wp-content/uploads/2017/06/Values_STEM-Report_20170403.pdf

Decker, A., & McGill, M. M. (2019, February). A topical review of evaluation instruments for computing education. In Proceedings of the 50th ACM Technical Symposium on Computer Science Education, Minneapolis, MN, USA (pp. 558–564).

Dorn, B., & Elliott Tew, A. (2015). Empirical validation and application of the computing attitudes survey. Computer Science Education, 25(1), 1–36. https://doi.org/10.1080/08993408.2015.1014142

Dorph, R., Bathgate, M. E., Schunn, C. D., & Cannady, M. A. (2018). When I grow up: The relationship of science learning activation to STEM career preferences. International Journal of Science Education, 40(9), 1034–1057. https://doi.org/10.1080/09500693.2017.1360532

Dorph, R., Cannady, M. A., & Schunn, C. D. (2016). How science learning activation enables success for youth in science learning experiences. The Electronic Journal for Research in Science & Mathematics Education, 20(8), 49–85.

Dweck, C. S. (2000). Self-theories: Their role in motivation, personality, and development. Psychology press.

Dweck, C. S. (2006). Mindset: The new psychology of success. Random House.

Dweck, C. S., & Leggett, E. L. (1988). A social-cognitive approach to motivation and personality. Psychological Review, 95(2), 256. https://doi.org/10.1037/0033-295X.95.2.256

Dweck, C. S., & Yeager, D. S. (2019). Mindsets: A view from two eras. Perspectives on Psychological Science, 14(3), 481–496. https://doi.org/10.1177/1745691618804166

Dweck, C. S., & Yeager, D. S. (2019). Mindsets: A view from two eras. Perspectives on Psychological Science, 14(3), 481–496.

Eccles, J. S., & Wigfield, A. (1995). In the mind of the actor: The structure of adolescents' achievement task values and expectancy-related beliefs. Personality & Social Psychology Bulletin, 21(3), 215–225. https://doi.org/10.1177/0146167295213003

Ferla, J., Valcke, M., & Cai, Y. (2009). Academic self-efficacy and academic self-concept: Reconsidering structural relationships. Learning and Individual Differences, 19(4), 499–505. https://doi.org/10. 1016/j.lindif.2009.05.004

Fields, D. A., Kafai, Y., Nakajima, T., Goode, J., & Margolis, J. (2018). Putting making into high school computer science classrooms: Promoting equity in teaching and learning with electronic textiles in exploring computer science. Equity & Excellence in Education, 51(1), 21–35. https://doi.org/10. 1080/10665684.2018.1436998

Flanigan, A. E., Peteranetz, M. S., Shell, D. F., & Soh, L. K. (2022). Shifting beliefs in computer science: Change in CS student mindsets. ACM Transactions on Computing Education (TOCE), 22(2), 1–24. https://doi.org/10.1145/3471574

Flapan, J., Ryoo, J. J., & Hadad, R. (2020, March). Building systemic capacity to scale and sustain equity in computer science through multi-stakeholder professional development. In 2020 Research on Equity and Sustained Participation in Engineering, Computing, and Technology (RESPECT), Portland, OR, USA (Vol. 1, pp. 1–8). IEEE.

Goode, J., Chapman, G., & Margolis, J. (2012). Beyond curriculum: The exploring computer science program. ACM Inroads, 3(2), 47–53. https://doi.org/10.1145/2189835.2189851

Goode, J., & Margolis, J. (2011). Exploring computer science: A case study of school reform. ACM Transactions on Computing Education, 11(2), 1–16. https://doi.org/10.1145/1993069.1993076

Gorson, J., & O'Rourke, E. (2019, July). How do students talk about intelligence? An investigation of motivation, self-efficacy, and mindsets in computer science. In Proceedings of the 2019 ACM Conference on International Computing Education Research, Toronto, ON, Canada (pp. 21–29).

Harel, I. E., & Papert, S. E. (1991). Constructionism. Ablex Publishing.Harter, S. (1999). The construction of the self: A developmental perspective. Guilford Press.

Kafai, Y. B., & Burke, Q. (2016). Connected gaming: What making video games can teach us about learning and literacy. MIT Press.

Kafai, Y. B., Fields, D. A., Lui, D. A., Walker, J. T., Shaw, M. S., Jayathirtha, G., Nakajima, T. M. . . . Goode, J., Giang, M. T. (2019, February). Stitching the loop with electronic textiles: Promoting equity in high school students' competencies and perceptions of computer science. In Proceedings of the 50th ACM technical symposium on computer science education, Minneapolis, MN, USA (pp. 1176–1182).

Kafai, Y. B., & Proctor, C. (2022). A revaluation of computational thinking in K–12 education: Moving toward computational literacies. Educational Researcher, 51(2), 146–151. https://doi.org/10.3102/ 0013189X211057904

Karwowski, M., & Kaufman, J. C. (Eds.). (2017). The creative self: Effect of beliefs, self-efficacy, mindset, and identity. Academic Press.

Kench, D., Hazelhurst, S., & Otulaja, F. (2016, July). Grit and growth mindset among high school students in a computer programming project: A mixed methods study. In Annual Conference of the Southern African Computer Lecturers' Association (pp. 187–194). Springer, Cham.

Kinnunen, P., & Simon, B. (2010, August). Experiencing programming assignments in CS1: The emotional toll. In Proceedings of the Sixth international workshop on Computing education research, Aarhus, Denmark. (pp. 77–86).

Lishinski, A., & Yadav, A. (2019). Motivation, Attitudes, and Dispositions. In S. Fincher & A. Robins (Eds.), The Cambridge Handbook of Computing Education Research (Cambridge Handbooks in Psychology, pp. 801-826). Cambridge: Cambridge University Press. https://doi.org/10.1017/ 9781108654555.029

Loksa, D., Ko, A. J., Jernigan, W., Oleson, A., Mendez, C. J., & Burnett, M. M. (2016, May). Programming, problem solving, and self-awareness: Effects of explicit guidance. In Proceedings of the 2016 CHI conference on human factors in computing systems, San Jose, CA, USA (pp. 1449–1461).

Lovell, E. (2014, July). Promoting constructive mindsets for overcoming failure in computer science education. In Proceedings of the tenth annual conference on International computing education research, Glasgow, United Kingdom (pp. 159–160).

Malmi, L., Sheard, J., Kinnunen, P., & Sinclair, J. (2020, August). Theories and models of emotions, attitudes, and self-efficacy in the context of programming education. In Proceedings of the 2020 ACM conference on international computing education research, Virtual Event, New Zealand (pp. 36–47).

Margolis, J. (2017). Stuck in the shallow end, updated edition: Education, race, and computing. MIT press.

Margolis, J., Goode, J., & Flapan, J. (2017). A Critical Crossroads for Computer Science for All: "Identifying Talent" or "Building Talent," and What Difference Does It Make? In Y. Rankin & J. Thomas (Eds.), Moving Students of Color from Consumers to Producers of Technology (pp. 1-23) IGI Global. https://doi.org/10.4018/978-1-5225-2005-4.ch001

Margulieux, L., Ketenci, T. A., & Decker, A. (2019). Review of measurements used in computing education research and suggestions for increasing standardization. Computer Science Education, 29(1), 49–78. https://doi.org/10.1080/08993408.2018.1562145

McGill, M. M., Decker, A., McKlin, T., & Haynie, K. (2019, February). A gap analysis of noncognitive constructs in evaluation instruments designed for computing education. In Proceedings of the 50th ACM Technical Symposium on Computer Science Education, Minneapolis, MN, USA (pp. 706– 712).

Morales-Navarro, L., Fields, D. A., Giang, M., & Kafai, Y. B. (2023). Designing bugs or doing another project: Effects on secondary students' self-beliefs in computer science. In P. Blikstein, A. Van Aalst, R. Kizito, & K. Brennan (Eds.). Proceedings of the 17th International Conference of the Learning Sciences - ICLS 2023. Montréal, Canada: International Society of the Learning Sciences.

Morales-Navarro, L., Fields, D. A., & Kafai, Y. B. (2021). Growing Mindsets: Debugging by Design to Promote Students' Growth Mindset Practices in Computer Science Class. In E. de Vries, Y. Hod, & J. Ahn (Eds.), Proceedings of the 15th International Conference of the Learning Sciences - ICLS 2021. (pp. 362–369). Bochum, Germany: International Society of the Learning Sciences.

Murphy, L., & Thomas, L. (2008, June). Dangers of a fixed mindset: Implications of self-theories research for computer science education. In Proceedings of the 13th annual conference on Innovation and technology in computer science education, Madrid, Spain (pp. 271–275).

Nolan, K., & Bergin, S. (2016, November). The role of anxiety when learning to program: A systematic review of the literature. In Proceedings of the 16th koli calling international conference on computing education research, Koli, Finland (pp. 61–70).

Oleson, A., Wortzman, B., & Ko, A. J. (2020). On the role of design in K-12 computing education. ACM Transactions on Computing Education (TOCE), 21(1), 1–34. https://doi.org/10.1145/3427594

Przybylla, M., & Romeike, R. (2014). Physical computing and its scope – Towards a constructionist computer science curriculum with physical computing. Informatics in Education, 13(2), 241–254. https://doi.org/10.15388/infedu.2014.14

Quille, K., & Bergin, S. (2020, June). Promoting a growth mindset in CS1: Does one size fit all? A pilot study. In Proceedings of the 2020 ACM Conference on Innovation and Technology in Computer Science Education, Trondheim, Norway (pp. 12–18).

Rangel, J. G. C., King, M., & Muldner, K. (2020). An incremental mindset intervention increases effort during programming activities but not performance. ACM Transactions on Computing Education (TOCE), 20(2), 1–18. https://doi.org/10.1145/3377427

Repenning, A., Webb, D. C., Koh, K. H., Nickerson, H., Miller, S. B., Brand, C., Her Many Horses, I. . . . Basawapatna, A., Gluck, F., Grover, R., Gutiérrez, K, Repenning, N. (2015). Scalable game design: A strategy to bring systemic computer science education to schools through game design and simulation creation. ACM Transactions on Computing Education (TOCE), 15(2), 1–31. https://doi.org/10.1145/2700517

Resnick, M., Maloney, J., Monroy-Hernández, A., Rusk, N., Eastmond, E., Brennan, K., Millner, A. . . . Rosenbaum, E., Silver, J., Silverman, B., Kafai, Y. B. (2009). Scratch: Programming for all. Communications of the ACM, 52(11), 60–67. https://doi.org/10.1145/1592761.1592779

Ryoo, J. J. (2019). Pedagogy that supports computer science for all. ACM Transactions on Computing Education (TOCE), 19(4), 1–23. https://doi.org/10.1145/3322210

Scott, M. J., & Ghinea, G. (2013). On the domain-specificity of mindsets: The relationship between aptitude beliefs and programming practice. IEEE Transactions on Education, 57(3), 169–174. https://doi.org/10.1109/TE.2013.2288700

Scott, M. J., & Ghinea, G. (2014, July). Measuring enrichment: The assembly and validation of an instrument to assess student self-beliefs in CS1. In Proceedings of the tenth annual conference on International computing education research, Glasgow, Scotland, UK (pp. 123–130).

Shaw, M. S., Fields, D. A., & Kafai, Y. B. (2019). Connecting with computer science: Electronic textile portfolios as ideational identity resources for high school students. International Journal of Multicultural Education, 21(1), 22–41. https://doi.org/10.18251/ijme.v21i1.1740

Smith, T. F., & Capuzzi, G. (2019). Using a mindset intervention to reduce anxiety in the statistics classroom. Psychology Learning & Teaching, 18(3), 326–336. https://doi.org/10.1177/ 1475725719836641

Steinhorst, P., Petersen, A., & Vahrenhold, J. (2020, August). Revisiting self-efficacy in introductory programming. In Proceedings of the 2020 ACM Conference on International Computing Education Research, Virtual Event, New Zealand (pp. 158–169).

Stout, J. G., & Blaney, J. M. (2017). "But it doesn't come naturally": How effort expenditure shapes the benefit of growth mindset on women's sense of intellectual belonging in computing. Computer Science Education, 27(3–4), 215–228. https://doi.org/10.1080/08993408.2018.1437115

Valentine, J. C., DuBois, D. L., & Cooper, H. (2004). The relation between self-beliefs and academic achievement: A meta-analytic review. Educational Psychologist, 39(2), 111–133. https://doi.org/10. 1207/s15326985ep3902_3

Vincent-ruz, P., & Schunn, C. D. (2017). The increasingly important role of science competency beliefs for science learning in girls. Journal of Research in Science Teaching, 54(6), 790–822. https://doi. org/10.1002/tea.21387

Waite, J., & Sentance, S. (2021). Teaching programming in schools: A review of approaches and strategies. Raspberry Pi Foundation.

Washington, A. N., Grays, S., & Dasmohapatra, S. (2016, June). The computer science attitude and identity survey (CSAIS): A novel tool for measuring the impact of ethnic identity in underrepresented computer science students. In 2016 ASEE Annual Conference & Exposition, New Orleans, Louisiana.

Weintrop, D., & Wilensky, U. (2019). Transitioning from introductory block-based and text-based environments to professional programming languages in high school computer science classrooms. Computers & Education, 142, 103646. https://doi.org/10.1016/j.compedu.2019.103646

Wigfield, A., & Meece, J. L. (1988). Math anxiety in elementary and secondary school students. Journal of Educational Psychology, 80(2), 210. https://doi.org/10.1037/0022-0663.80.2.210

Woods, D. M. (2020). Using goal setting assignments to promote a growth mindset in IT students. Information Systems Education Journal, 18(4), 4–11.

Wylie, R. C. (1979). The self concept. In Volume 2: Theory and research on selected topics. University of Nebraska Press.

Yeager, D. S., & Dweck, C. S. (2012). Mindsets that promote resilience: When students believe that personal characteristics can be developed. Educational Psychologist, 47(4), 302–314. https://doi. org/10.1080/00461520.2012.722805

# Appendix 1.

| Construct | CSIBI Items and Scales (with constructs and construct items presented to students in randomized order) | | | | |
|---|---|---|---|---|---|
| Problem Solving Competency Beliefs | The following questions ask about your perspectives towards specific computer science activities. Please indicate how much you agree or disagree with the following statements: <br><br> I think I am very good at: | Strongly Disagree | Disagree | Agree | Strongly Agree |
| | Figuring out how to fix things that don't work. | ○ | ○ | ○ | ○ |
| | Explaining my solutions to technical problems. | ○ | ○ | ○ | ○ |
| | Solving problems. | ○ | ○ | ○ | ○ |
| | Coming up with new ways to solve technical problems. | ○ | ○ | ○ | ○ |
| | Coming up with new ideas when working on projects. | ○ | ○ | ○ | ○ |
| Fascination in Design | The following questions ask about your perspectives towards specific computer science activities. Please indicate how much you agree or disagree with the following statements: | Strongly Disagree | Disagree | Agree | Strongly Agree |
| | I love designing things! | ○ | ○ | ○ | ○ |
| | Designing new things makes me feel excited. | ○ | ○ | ○ | ○ |
| | I talk about how things work with friends or family. | ○ | ○ | ○ | ○ |
| Value of CS | The following questions ask about your perspectives towards the value of computer science. Please indicate how much you agree or disagree with the following statements: | Strongly Disagree | Disagree | Agree | Strongly Agree |
| | Knowing computer science is important for contributing to my community. | ○ | ○ | ○ | ○ |
| | Knowing computer science is important for me in the future. | ○ | ○ | ○ | ○ |
| | I want to learn as much as possible about computer science. | ○ | ○ | ○ | ○ |
| | Complete the statement so that it reflects your personal opinion: | None of my classes | Few of my classes | Most of my classes | All of my classes |
| | Thinking like a computer scientist will help me do well in… | ○ | ○ | ○ | ○ |
| CS Creative Expression | The following questions ask about your perspectives towards specific computer science activities. Please indicate how much you agree or disagree with the following statements: | Strongly Disagree | Disagree | Agree | Strongly Agree |
| | I can be creative in computer science. | ○ | ○ | ○ | ○ |
| | I can express myself in computer science. | ○ | ○ | ○ | ○ |
| | I can make things that are interesting to me in computer science. | ○ | ○ | ○ | ○ |
| E-textiles Coding Self-efficacy | The following questions ask about your perspectives towards specific computer science activities. Please indicate how much you agree or disagree with the following statements: | Strongly Disagree | Disagree | Agree | Strongly Agree |
| | I am confident that I can understand Arduino errors | ○ | ○ | ○ | ○ |

| | (e.g., was not declared in this scope, expected ';' before). | | | | |
|---|---|---|---|---|---|
| | I am confident I can write code for a simple e-textiles project. | ○ | ○ | ○ | ○ |
| | I am confident I can create a functional program for an e-textiles project that uses both sensors and actuators (e.g., LEDs, speakers). | ○ | ○ | ○ | ○ |

| | The following questions ask about your perspectives towards programming. Please indicate how much you agree or disagree with the following statements: | | | | |
|---|---|---|---|---|---|
| | | Strongly Disagree | Disagree | Agree | Strongly Agree |
| **Programming Mindset (Fixed and Growth presented together)** | I have a fixed level of programming ability, and not much can be done to change it. | ○ | ○ | ○ | ○ |
| | I can learn new things about software development, but I cannot change my basic ability for programming. | ○ | ○ | ○ | ○ |
| | To be honest, I do not think I can really change my ability for programming. | ○ | ○ | ○ | ○ |
| | No matter who you are, you can significantly change your programming ability. | ○ | ○ | ○ | ○ |
| | I can always substantially change my programming ability. | ○ | ○ | ○ | ○ |
| | No matter how much programming ability I have, I can always change it quite a bit. | ○ | ○ | ○ | ○ |
| | I can change even my basic programming ability considerably. | ○ | ○ | ○ | ○ |

| | The following questions ask about your perspectives towards programming. Please indicate how much you agree or disagree with the following statements: | | | | |
|---|---|---|---|---|---|
| | | Strongly Disagree | Disagree | Agree | Strongly Agree |
| **Programming Anxiety** | I often worry that it will be difficult for me to complete debugging exercises. | ○ | ○ | ○ | ○ |
| | I often get tense when I have to debug a program. | ○ | ○ | ○ | ○ |
| | I get nervous when trying to solve programming bugs. | ○ | ○ | ○ | ○ |
| | I feel helpless when trying to solve programming bugs. | ○ | ○ | ○ | ○ |

| | The following questions ask about your perspectives towards programming. Please indicate how much you agree or disagree with the following statements: | | | | |
|---|---|---|---|---|---|
| | | Strongly Disagree | Disagree | Agree | Strongly Agree |
| **Programming Self-concept** | I learn programming quickly. | ○ | ○ | ○ | ○ |
| | I have always believed that programming is one of my best subjects. | ○ | ○ | ○ | ○ |
| | In my programming classes, I can solve even the most challenging problems. | ○ | ○ | ○ | ○ |

| **Demographic Questions (always requested at *end* of the survey)** | |
|---|---|
| **Previous experience with computing** | Before this computer science class, did you take any computer science classes or workshops? Yes \| No |
| **Gender** | Please indicate your gender: Female \| Male \| Non-binary \| Other \| Decline to indicate |
| **Home Language** | How often do people in your home talk to each other in a language other than English? Never \| Once in a while \| About half of the time \| Most of the time \| All the time |
| **Family College Attendance History** | Who in your immediate family has attended college? (Select all that apply): Mother \| Father \| Sibling \| Grandparent \| Other (please specify) \| No one |
| **Race & Ethnicity*** | Do you identify as Latina, Latino, Latinx or Hispanic? Yes \| No<br>Please indicate your race (Select all that apply): African American / Black \| Asian American / Pacific Islander \| Native American \| White / Caucasian \| Other (please specify) \| Decline to indicate |

* **Note:** Not included in this study to protect participant's privacy but encouraged for future studies.